\documentstyle[12pt]{article}
\def\a{\begin{eqnarray}}
\def\b{\end{eqnarray}}
\def\0{\nonumber}
\def\ba{\begin{array}}
\def\ea{\end{array}}

\def\cL{{\cal L}}
\begin{document}
\begin{titlepage}
\begin{center}
\hfill    SISSA 130/97/EP/FM\\
\vskip 5pt
\hfill    hep-th/9710063\\
\vskip 30pt

{\Large \bf Perturbative Anomalies of the M--5--brane}
\vskip 30pt

{ \large L. Bonora$^{a,b}$,~ C.S. Chu $^{a}$ and M. Rinaldi $^{c\dagger}$}

\vskip 30pt

{\em ~$~^{(a)}$ International School for Advanced Studies (SISSA/ISAS),}\\
{\em Via Beirut 2, 34014 Trieste, Italy}\\
\vskip 5pt
{\em ~$~^{(b)}$  INFN, Sezione di Trieste, Italy}\\
\vskip 5pt
{\em ~$~^{(c)}$  Dipartimento di Matematica, Universit\'a di Trieste, }\\
{\em P.le Europa 1, 34127 Trieste, Italy}

\vspace{60pt}

{\bf Abstract}

\end{center}

We discuss several mechanisms to cancel the anomalies of a 5--brane
embedded in M--theory. Two of them work, provided we impose suitable
conditions either on the 11--dimensional manifold of M--theory or on 
the 4--form field strength of M--theory.

{
\vfill\leftline{} \vfill
\vskip  10pt
\footnoterule
\footnotesize $~^{\dagger}$
E-mail: bonora@sissa.it, cschu@sissa.it, rinaldi@uts.univ.trieste.it
\vskip -12pt
}
\vskip  10pt

\end{titlepage}

\section{Introduction}
The world--volume massless content of the M theory 5--brane fills up the tensor 
multiplet of $(2,0)$ supersymmetry in 6D. It contains 8 chiral spinors, 5 
bosons which represent the directions transverse to the 5--brane and 3 
components of an antiselfdual two--form. So the corresponding theory is 
potentially anomalous. The actual calculation shows that the chiral anomaly
always vanishes when the 5--brane is considered in isolation, but does not 
in general vanish when the 5--brane is embedded into the M theory. 
Geometrically
this corresponds to having a six--dimensional manifold $W$, the world--volume 
of the 5--brane, embedded in the eleven--dimensional manifold $Q$ of M theory,
(we are closely following the analysis and notations of \cite{witten1}). 

The anomalies in question arise from a breakdown of invariance under the 
diffeomorphisms of $Q$ that preserve the embedding of $W$ in $Q$. More 
precisely,
the tangent bundle $TQ$ of $Q$ restricted to $W$ decomposes according to
$TQ|_W = TW \oplus N$, where $TW$ and $N$ are the tangent bundle and the
normal bundle of $W$, respectively. A Riemannian metric on $Q$ induces a 
Riemannian metric on $W$ and a metric and an $SO(5)$ connection on $N$. The 
relevant diffeomorphisms of $Q$ are those that map $W$ to $W$. Any such 
diffeomorphism generates a diffeomorphism of $W$; if the diffeomorphism
induced on $W$ is the identity, it determines a gauge transformation in
$N$. Therefore the anomalies to be considered are the usual gravity anomalies
in $TW$ and the gauge anomalies in $N$. Their evaluation has been carried out
in \cite{witten1}: the contributions from the fermions, the chiral two--form
and the inflow from the bulk do not cancel completely, and the overall chiral 
anomaly has a remarkably simple expression: it is generated via the usual
descent equations from the 8--form
\a
&&{1\over 24} p_2(N)=  d\omega_7(N),\0\\
&&0=\delta \omega_7(N) +d {\cal A}_6\label{p2N},
\b 
where $p_2(N)$ is the second Pontrjagin class of the normal bundle of the 
5--brane world--volume and, for later use, we have introduced the corresponding
properly normalized Chern--Simons term and 6--form anomaly $\omega_7(N)$ 
and ${\cal A}_6$,
respectively. Here $\delta$ represents the gauge transformations induced on the
normal bundle.

It is shown in \cite{witten1} that this residual anomaly can be canceled 
by means of a suitable 6D counterterm in the case in which $Q$ is the 
product of a ten dimensional manifold and a circle, i.e. when M--theory reduces 
to type IIA superstring theory (in the limit of small radius) and the 5--brane 
reduces to the solitonic 
5--brane of type IIA. Here we want to discuss the problem in general. 
The result of our investigation is that there are mechanisms
to cancel the residual anomaly. However they work only at some price, which 
consists of imposing some conditions either on the topology of $Q$ or 
on $F_4$, the 4--form field strength of M--theory.
In section 2, we determine two such conditions.
Starting from the discussion of a topological counterterm, we show how to
characterize $Q$ so that this counterterm be well defined. We can alternatively
construct a world--volume counterterm that cancels the anomaly, using
the fact that a 5--brane is {\it framed}, see below. This works provided
an additional condition on $F_4$ is imposed. An important problem, which we 
do not discuss in this paper, concerns the physical (or unphysical) nature of 
the framing.

To complete our discussion, in section 3 we have collected other cancellation 
mechanisms which, in our opinion, are not admissible or do not work. 

Before we start our analysis, let us recall a few basic facts about anomalies.
An anomaly, being the result of a local calculation, must be represented 
by a global basic form, therefore in 
eq.(\ref{p2N}), $\omega_7(N)$ must also be a basic form.  
A global basic form $\omega_7(N)$ in $W$ can be written by introducing 
a reference connection. Let $A$ be the generic connection (with curvature $F$)
in the principal bundle $P(W, SO(5))$ to which the normal bundle
is associated, and $A_0$ the reference connection. Then 
\a
 \omega_7(N)&=& 
4\int_0^1dt~ {\rm P}_4 (A-A_0, {\cal F}_t,{\cal F}_t,{\cal F}_t)
\0\\
&=&CS(A)-CS(A_0) + d S(A,A_0),\label{CS1}
\b
where   
\a
CS(A) &=& 4\int_0^1 dt~ {\rm P}_4(A,F_t,F_t,F_t)\label{CS2}
\b
and  $S(A,A_0)$ is a
suitable 6--form in $P$ (see \cite{chern}).
Here ${\cal F}_t$ is the curvature of $tA+(1-t)A_0$
and $F_t= tdA +t^2/2[A,A]$.
${\rm P}_4$ is the fourth
order symmetric $SO(5)$--ad--invariant polynomial corresponding to $p_2(N)/24$,
\a
\frac{1}{24}p_2(N) = {\rm P_4} (F^4).\label{P4}
\b
The three forms
in the RHS of (\ref{CS1}) are all defined in the total space of $P(W, SO(5))$,
while the LHS is basic. 

We remark in particular that $\omega_7$ being basic means that it 
vanishes identically in $W$, as $W$ is a 6--dimensional space. This fact is 
immaterial as long as we deal with descent equations (for example, in
section 2 of \cite{BCRS1} it is shown how to reconcile it with the descent
equations), but becomes very important if 
one wants to use $\omega_7(N)$ to construct an action term. This will be one 
of our main concerns in this paper.

\section{Mechanisms for the anomaly cancellation}

Let us start by recalling that a 5--brane magnetically couples to M--theory
via the equation, \cite{witten2},
\a
d F_4 = \delta_W. \label{mag}
\b
where $F_4$ is the 4--form field strength of M--theory, and $\delta_W$
is a closed form which represents the Poincar\`e dual of the world--volume
manifold $W$ (see also below). As long as (\ref{mag}) is a defining 
equation for the 5--brane, it implies, by Poincar\`e duality, that 
$W$ is the boundary of some seven--dimensional manifold $Y$ in $Q$,
\a
W=\partial Y. \label{Y}
\b

In the following we are going to exploit this fact in order to cancel the
residual anomaly. We are going to study two mechanisms.

\subsection{A topological counterterm}

We examine the possibility to cancel the anomaly of the M--5--brane 
by means of the counterterm,
\a
S_{top} = \int_Y \omega_7 .\label{WZ}
\b
where eq.(\ref{Y}) is understood.
Formally,
\a
\delta S_{top} = -\int_Y d{\cal A}_6 = - \int_W {\cal A}_6.\0
\b

However we must exert some care in order for this to make sense.
We have just recalled that $\omega_7(N)$, as defined in (\ref{p2N}), is a basic 
form in $W$, and so it vanishes identically. Therefore
in order to have a non--trivial counterterm, we must extend $\omega_7(N)$
to $Y$ in a non--canonical (constructive) way, because the only canonical 
extension of 0 is 0; this in turn requires that we extend the connection and,
before that, the normal bundle.

In summary, we must make sure that:
\begin{itemize}
  
\item [1.] the normal bundle over $W$ extends to a bundle over $Y$;
\item [2.] the connection on the normal bundle extends too;
\item [3.] the counterterm (\ref{WZ}) is independent of the choice of  $Y$.
\end{itemize}

Conditions 1 and 2 are needed to ensure that the counterterm (\ref{WZ})
makes sense. Let us analyze them first.

To this end the normal bundles need a more precise notation: 
$N(X,Z)$ will denote the normal bundle of $X$ embedded in $Z$. For instance, 
according to this new notation, $N\equiv N(W,Q)$.  
First we have 
\a
i^* T Q\equiv  T Q|_W=  T W\oplus N(W,Q), \label{norm1}
\b
where $i\colon W  \to Q$ is the embedding and 
\a
 T Q|_W =  T Y|_W \oplus N(Y,Q).\0
\b
As for $Y$, it is not closed, still we have
\a
  T Y|_{W}= T W\oplus N(W,Y) .\0
\b
It follows that 
\a
N(W,Q)=N(W,Y)\oplus N(Y,Q)|_W .\0
\b
Notice that $L \equiv N(W,Y)$ is a one-dimensional bundle. 
Since it is orientable, it is trivial and so we can write 
\a
N=N'\oplus L, \quad N'\equiv N(Y,Q)|_W, \quad L= W\times {\bf R},\label{decomp}
\b
where $N'$ is an $SO(4)$-bundle, i.e. the gauge group $SO(5)$ reduces to 
$SO(4)$. The decomposition (\ref{decomp}) implies in particular that 
$p_2(N)=p_2(N')$.   
Since 
\a 
p_2(N')=p_2(N(Y,Q))|_W\0, 
\b   
$p_2(N)$ extends to $0\in H^8(Y)$. As forms we can write
\a
p_2(N)=e(N(Y,Q))|_{W}^2, \0
\b  
where $e$ denotes the Euler class.

After these preliminaries let us discuss the conditions 1 and 2 above.
Using (\ref{decomp}), one can trivially extend $L$ and $N'$ to bundles over 
$Y$ as follows:
\a
&&N'=N(Y,Q)|_W \rightarrow N(Y,Q), \0\\ 
&&L =W\times {\bf R}  \rightarrow \tilde L = Y\times {\bf R}. \0
\b
Consequently $N$ extends in a natural way as
\a
N\rightarrow \tilde N = N(Y,Q) \oplus \tilde L. \label{ext}
\b
We can now extend the connection over this bundle as follows. Let us construct
a connection over $\tilde N$ by taking the trivial connection in $\tilde L$
plus the connection induced from $Q$ in $N(Y,Q)$. This is of course an
extension of the connection induced from $Q$ in $N$.
Therefore also the form $\omega_7$ extends, as well as all the operations on it.
All this is made possible by (\ref{decomp}).

It remains for us to discuss condition 3. By a standard argument the term
(\ref{WZ}) will not depend on the 
particular $Y$ 
if 
\a
{1\over {2\pi}} \int_X \omega_7 \in {\bf Z}\label{integer}
\b
for any compact 7--manifold $X$ without boundary.
Looking at (\ref{CS1}) it is actually easy to make a more precise statement
about the integral in (\ref{integer}). We remark that $\omega_7$ vanishes
when $A=A_0$. Now, a generic connection $A$ can be continously joined to $A_0$.
Therefore also the results of the integral in (\ref{integer}) when evaluated
at $A_0$ and at $A$ should be continuously connected. This means that
the only value of the integral compatible with (\ref{integer}) is {\it zero}.

This seems to be a very strong condition on the dynamics of the 5--brane,
anyhow a hard to interpret one. Therefore we are prompted to take a 
different 
attitude: we consider for the time being unavoidable the idea of the action 
term (\ref{WZ}) being dependent on $Y$ \footnote{We would like to point out that
a similar problem of $Y$ dependence appears in the attempts at writing
a $\kappa$--symmetric action of the 5--brane in M--theory \cite{schwarz,pst}. 
Perhaps the two problems are connected.}. With this attitude, 
a 5--brane in M--theory is characterized not only by the usual data (in
particular by eq.(\ref{mag})), but also by specifying a seven--manifold $Y$
that bounds its world--volume. We hope a better understanding of M--theory, 
when available, will permit us to decide whether this $Y$--dependence is a
physical characteristics of the 5--brane or a mere `gauge' freedom.

However, if the 5--brane is to be specified 
by some other data beside the traditional ones, there is a less stringent
and more manageable way to do it than by specifying a bounding manifold 
$Y$: we can simply add the specification of a `collar' or `framing' 
attached to the 5--brane.
This procedure will be illustrated in the following subsection.

\subsection{A world--volume counterterm}

If the integral in (\ref{integer}) vanishes, it
ensures the anomaly cancellation for any 5--brane embedded in $Q$.
In this subsection we would like to explore a different point of view. We
study anomaly cancellation conditions for a single 5--brane (not for all
possible 5--branes) embedded in $Q$.
The previous analysis told us that any 5--brane
is {\it framed}, i.e. that its normal bundle splits, as in (\ref{decomp})
-- with rather non--standard terminology we call such a splitting a framing.
If we consider a 5--brane with a {\it fixed} framing, we can construct a 
counterterm, akin to the one in \cite{witten1},
which, at some additional cost, cancels the anomaly.
We stress again that the fixed framing in this section is part of 
the definition of the 5--brane and is a local property as opposed to the
global existence of a bounding seven--manifold $Y$.

Therefore, let the normal bundle $N$ split as
\a
N=N' \oplus L, \label{C1}
\b
where $N'$ is an $SO(4)$-bundle and $L$ is  a trivial line bundle.
We have 
\a
p_2(N)=p_2(N') =e(N')^2 \label{pp1}
\b
and
\a
\Phi(N)=pr_1^*\Phi(N')\wedge pr_2^*\Phi(L),
\label{pp2}
\b
where $pr_i$, $i=1,2$ are the projections in the decomposition (\ref{C1})
of $N$ and $\Phi(E)$ is the Thom class \cite{bott} of a real 
vector bundle $E$ over $W$. $\Phi$ and $e$ are related by 
\a
\sigma_0^* \Phi(N') = e(N'),
\b 
where $\sigma_0\colon W\to N' $ is the zero section of the bundle $N'$. From
now on, for simplicity, we will drop the pull--back symbols.

Let $N_\epsilon(W)\subset N$ be a suitable tubular
neighborhood of $W$ of size $\epsilon$ embedded in $Q$.
If we identify $N_\epsilon(W)$ with the corresponding submanifold of $Q$,
then since  the support of $\Phi(N)$ can be taken to lie inside $N_\epsilon(W)$, 
one can view $\Phi(N)$ as a form 
on $N_\epsilon(W)$ and identify it with the Poincar\'e dual  of
$W$, represented  by the form $\delta_W$.

Let $\omega_3$ be the basic 3--form constructed out of the polynomial
corresponding to $1/24 e(N')$ just like $\omega_7$ in eq.(\ref{CS1}).
Using the descent equations we get 
\a
\delta \omega_3 + d a=0, \0
\b
(\ref{pp1}) says that the integrated anomaly for the framed 
5--brane is given by 
\a
\int_W{\cal A}_6 = \int_W e' \wedge a.\0
\b
with $e' \equiv e(N')$. 

Let $v$ be  a nonvanishing vertical vector field on $L$ 
such that its contraction with $\Phi(L)$ satisfies
\a
i_v \Phi(L)|_W =1. \label{v}
\b
The existence of such a field follows from the 5--brane being framed.
If $s$ is the coordinate along the fiber of $L$, we can simply choose
$v =\alpha \frac{\partial}{\partial s}$ with the constant $\alpha$ such that
(\ref{v}) be satisfied. 
If we now assume that the Lie derivative of $F_4$ in the $L$-direction
is zero on $W$,
\a
{\cL}_v F_4|_W =0.\label{C2}
\b
then, using (\ref{mag}), (\ref{pp2}), (\ref{v}) and (\ref{C2}),
it is easy to see that
\a
d i_v F_4|_W = - e' .\label{df4e}
\b
Now, consider the counterterm
\a
S= \int_W i_v  F_4\wedge \omega_3.\label{counter}
\b
Its gauge variation is, using (\ref{df4e}),
\a
\delta S= -\int_W  i_v F_4\wedge d a =
\int_W d i_v  F_4\wedge a = -\int_W e'\wedge a.\0
\b  
Hence the counterterm cancels the anomaly.

The case of IIA 5--brane, in which $Q=M\times S^1$, follows as a particular 
situation of the above. In this case the framing is determined by the 
background geometry.
 
Eq.(\ref{C2}), i.e. the constancy of $F_4$ along $L$ on $W$,
is the only condition required for this mechanism to work. It must be added to 
the defining eq.(\ref{mag}). We remark that the counterterm (\ref{counter})
depends on the vector field $v$, i.e. on the framing. In this section we have 
picked a definite one. However a 5--brane can have in general several distinct 
framings. In the previous subsection we saw that the 5--brane can have several
distinct $Y$'s that bound it. Framings are local objects on the 5--brane,
while the $Y$'s are non--local, and, if we cannot get rid of the $Y$
dependence in (\ref{WZ}), this may be a good reason to prefer the mechanism 
outlined in this subsection. Let us repeat again that the anomaly cancellation
problem does not end here. The question of the physicality of $v$ (or $Y$) 
remains open. To understand it a more complete formulation of $M$--theory with 
5--branes would be extremely helpful. 

\section{Other mechanisms}

\subsection{A non--minimal 5--brane}

The expression $p_2(N)/24$ is obtained via index theorem by summing the
contribution from the chiral world--sheet fermions and antiself--dual 
two--form to the inflow anomaly. One possibility to cancel the anomaly is to
consider a non--minimal 5--brane, i.e. to add $N=2$ supermultiplets.
These can be \cite{strat}:
\begin{itemize}
\item {\it T}=tensor ~~(3,1;1) + (1,1;5)+(2,1;4)
\item {\it G}=gravity ~(3,3;1)+(1,3;5)+(2,3;4)
\item {\it V}=vector ~~(3,2;1)+(1,2;5)+(2,2;4)
\end{itemize}
where the first two entries label the representations of the little group
$SO(4)\sim SU(2)\times SU(2)$ and the third entry labels the representations
of $USp(4)\sim SO(5)$, which we identify with the structure group of the
normal bundle.

Then we use the relevant index theorems, 
\a
&&I_{(3,1;1)} = -{1\over 8}L(TW), \0\\
&&I_{(2,1;4)} = {1\over 2}\hat A(TW) ch S_4(N), \0\\
&&I_{(1,3;5)} = {1\over 8} L(M) ch_2S_5(N),\0\\
&&I_{(2,3;4)} = - {1\over 2} \hat A(TW)(tr (e^{{i\over 2\pi}R})-1)chS_4(N),\0\\
&&I_{(3,2;1)} = {1\over 2} \hat A(TW)(tr (e^{{i\over 2\pi}R})-1),\0\\
&&I_{(1,2;5)} = -{1\over 2}\hat A(TW) ch S_5(N), \label{index}
\b
where $L(M)$ is the Hirzebruch polynomial, $\hat A(TW)$ the arithmetic genus
of the tangent bundle, $chV$ denotes the Chern character of the vector bundle 
$V$, $R$ is the curvature of $W$, $S_n(N)$ denotes the spin bundle tensored
with the vector bundle associated to the normal bundle via the representation 
${\bf n}$ of $SO(5)$ with $n=4,5$.
Finally $ch_2 V= \sum_{k=0}^\infty2^k ch^{(k)}V$,
where $ch^{(k)}V$ is the $k$-th order term in $chV$, see for example \cite{law}.

We write the above contributions in terms of Pontrjagin classes $p_i(TW)$ and
$p_i(N)$. For our purposes it will be enough to consider 
the irreducible contribution in 
terms of second Pontrjagin classes only. For the various multiplets they are 
as follows
\a
&&T:\quad\quad {1\over 48} (p_2(N)-p_2(TW))\0\\
&&G:\quad\quad -{21\over 48}(p_2(N)-p_2(TW))\0\\
&&V:\quad\quad {4\over{48}}(p_2(N)-p_2(TW))\label{p2}
\b
Since the corresponding inflow anomaly contribution is proportional to
$p_2(N) + p_2(TW)$, it is evident that by adding new multiplets it is not
possible to cancel the anomaly
\footnote{The presence of $G$ in the theory 
introduces anyhow new world--volume anomalies.}. 

\subsection{A generalized Green--Schwarz mechanism}

A six--form potential $C_6$ couples naturally to the world--volume of the 
5--brane. One therefore may wonder whether the term $\int_{W} C_6$ in the 
action could cancel the anomaly provided
\a
\delta \int_W C_6 = -\int_W{\cal A}_6. \label{desc6}
\b
Following the analogy with the original Green--Schwarz mechanism, 
eq.(\ref{desc6}) should come from
\a
&&dF_7 = 1/24 p_2(N), \label{wrong1}\\
&&F_7 - dC_6 = \omega_7(N), \label{wrong2}
\b
where the first equation is the modified Bianchi identity for the gauge
invariant curvature $F_7$ of $C_6$. However this fails in two respects. 

First, the RHS of the above two equations vanishes.
They are to be understood as equations in $Q$, but $p_2(N)$ and $\omega_7(N)$
are basic forms in $W$ and thus vanish identically. The statement concerning
$p_2(N)$ is obvious; as for $\omega_7(N)$ we have already explained this point
in the introduction. The original Green--Schwarz mechanism involved 2--, 3-- 
and 4--forms in a 10 dimensional space, and therefore the same conclusion was 
prevented. In conclusion, the transformation (\ref{desc6}) does not follow from 
the Bianchi identity. One might try to extend $\omega_7$ to a non--vanishing
7--form in $Q$. However, as we explained in section 2, there may
be many non--trivial extensions, but they are non--canonical and one has 
to specify them in a constructive manner, because any extension provides 
presumably different conditions either on $Q$ or on the 5--brane embedding. 
An example of extension has been given in section 2.1 and we could repeat 
here a similar analysis. However it would be an academic exercise since there
is a second objection.

The second reason is that in M--theory in the presence of a 5--brane, 
eq.(\ref{wrong1}) is not quite correct, but must be modified, \cite{witten2}. 
As an equation of motion of M--theory coupled to
a 5--brane, it should contain other terms:
\a
dF_7 + {1\over 2}F_4\wedge F_4 - T \wedge \delta_W + I_8 =0, \label{wright1}
\b
where $F_7$ is the Hodge dual of $F_4$, $T$
is the anti--self--dual 3--form on $W$ satisfying \cite{town} $dT =F_4|_W$ 
and is suitably extended to $Q$. 
$I_8$ is the 8--form responsible for the inflow anomaly.

Eq.(\ref{wright1}) is consistent, but we do not see how eq.(\ref{desc6})
can be derived from it. For this reason we discard also this cancellation 
mechanism.

\subsection{A more exotic possibility}

The analysis below grew out of an attempt to generalize a suggestion 
contained in \cite{witten1}. 
Let us start with writing the residual anomaly in a form suitable
for our discussion. To this end we introduce the $4\times 4$ $SO(5)$ gamma 
matrices, $\gamma_a, ~a=1,...,5$, which satisfy
\a
\{\gamma_a, \gamma_b\} = 2 \eta_{ab},\qquad\quad \eta_{ab} = 
{\rm diag}(1,1,1,1,1). \label{gamma}
\b
In particular we have
\a
{\rm Tr}(\gamma_a \gamma_b\gamma_c \gamma_d \gamma_e) = 4 \epsilon_{abcde},
\qquad \epsilon_{12345}=1. \label{epsilon}
\b
It is well--known that the generators of $SO(5)$, in the representation {\bf 4},
can be written as
\a
\Sigma_{ab} = - {1\over 4}[\gamma_a,\gamma_b]. \label{sigma}
\b
We write the connection and curvature in the normal bundle $N$ as 
$A = A^{ab}\Sigma_{ab}$ and $F= F^{ab}\Sigma_{ab}$. Using this we find
\a
p_2(N) = - \frac{1}{(2\pi)^4 4!} \sum_{a=1}^5 Tr(\gamma_a FF)Tr(\gamma^a FF),\0
\b
where the trace is over the gamma matrices, i.e. over the representation 
${\bf 4}$ of $SO(5)$. This is in explicit form the decomposition proposed 
in \cite{witten1}.
However we have not been able to implement the descent equations on a single
factor $\chi_a = Tr(\gamma_a FF)$. For example, it is easy to see that there
does not exist any combination $\phi_a$ of three forms $Tr(\gamma_a XY)$, 
where $X,Y$ are either $A,dA$ or $[A,A]$, such that 
\a
\chi_a = d\phi_a + w A_{a}{}^b \phi_b\0
\b 
for any weight $w$.

Trying to identify a non--abelian cancellation mechanism, by generalizing the
above ideas, we were lead to a Green--Schwarz scheme, but, this time, with
a non--abelian two--form potential. Non--Abelian tensor potentials have 
appeared recently in the literature, \cite{douglas,hull}.
What follows will be very sketchy, we only present an example of the type 
of arguments we have used to exclude this mechanism too.

One can extract the explicit form of the anomaly
from the eight--form (\ref{P4}) via the descent equations and get
\a
{\cal A}_6  = 
6 \int_0^1 dt ~(t-1)~{\rm P_4} (dc, A, F_t,F_t), \label{anom}
\b
where $c = c^{ab}\Sigma_{ab}$ is the gauge parameter.

To cancel this anomaly one may think of the following counterterm
\a
{\cal C} = \int_0^1dt \Big\{ f ~{\rm P_4}({\cal B}, F_t,F_t) 
+g ~{\rm P_4}({\cal B}, (F_t,F_t)')
+ h~ {\rm P_4}({\cal B}, (F_t,F_t)'')\Big\} .\label{count}
\b
Here $f,g,h$ are $t$--dependent function to be determined and $'$ denotes a 
derivative with respect to $t$. The potential ${\cal B}$ is a two--form 
that belongs to the symmetric part of ${\bf 10}\otimes {\bf 10}$ 
decomposition,
\a
{\bf 10}\otimes {\bf 10} = {\bf 35_s + 35_a' +14_s + 10_a + 5_s +1_s}, 
\label{tprod}
\b
i.e. it is a tensor ${\cal B}^{ab,cd}$ antisymmetric in 
$a \leftrightarrow b$ and 
$c \leftrightarrow  d$ and symmetric in the exchange of $ab$ with $cd$.
A gauge transformation, $\delta A = dc + [A,c]$, is assumed to act on
${\cal B}$ for example as follows
\a
\delta {\cal B} = [{\cal B},c] + \beta dc A. \label{gaugetr}
\b
In the last term we understand that we take the symmetric part of the tensor
product between the Lie algebra generators. $\beta$ is another $t$--dependent
function to be determined. This implies in particular that ${\cal B}$ is 
also $t$--dependent. 

We now impose that
\a
\delta {\cal C} = 
-\int_0^1 dt ~(t-1)~{\rm P_4} (dc ,A, F_t,F_t) \0
\b
and work out the conditions for which this is satisfied. They consist
of differential and algebraic equations on $f,g,h, \beta$, plus boundary 
conditions at $t=0$ and $t=1$. The system does not admit non--trivial solutions.

\section{Comments}

In section 2 we have presented cancellation schemes for the residual anomaly
of the 5--brane embedded in M--theory. They work provided certain conditions are
imposed either on the topology of $Q$ or on the dynamics of the 5--brane. 
These mechanisms may not be the only ones.
For example, one could try a counterterm 
\footnote{This counterterm is also studied in ref.\cite{alwis}. We
became aware of this reference after completion of our paper. We thank 
P.West for pointing it out to us. Section 7 of \cite{alwis} also analyzes the
5--brane anomaly problem in M--theory; but there different cancellation 
mechanisms are considered, except for one case, corresponding to our 
section 3.3, where our conclusion is anyhow different.}
\a
\int_Q F_4\wedge \omega_7. \label{f4}
\b
This term identically vanishes for the reasons explained in subsection 2.1
and 3.2. However one may try to extend $\omega_7$ to a non--vanishing 
7--form in $Q$. Our point of view should be clear by now:
as explained above, this extension, if it exists, is non--canonical and must 
be explicitly constructed.  
\vskip.3cm

{\Large \bf Acknowledgements}

We would like to thank L. Baulieu, P. Pasti, E. Sezgin, J. Schwarz,
A. Schwimmer, M. Tonin and P. West for discussions. This research was partially 
supported by EC TMR Programme, grant FMRX-CT96-0012.

\end{document}